\documentclass[aps,prb,twocolumn,superscriptaddress,showpacs,showkeys]{revtex4}
\usepackage{bm,amsmath,graphicx,dcolumn}
\begin{document}

\title{On the high accuracy lattice parameters determination by n-beam diffraction: Theory and application to the 
InAs quantum dots grown over GaAs(001) substrate system}
\author{L.H. Avanci}
\email{lhavanci@if.usp.br}
\affiliation{Instituto de F\'{\i}sica, Universidade de S\~ao Paulo, CP 66318, 05315-970 S\~aoPaulo, SP, Brazil}
\author{C.M.R. Rem\'edios}
\email{rocha@fisica.ufc.br}
\affiliation{Departamento de F\'{i}sica, Universidade Federal do Cear\'a, Campus do Pici, CP 6030, 60455-760, Fortaleza, CE, Brazil}
\author{A.A. Quivy}
\email{aquivy@macbeth.if.usp.br}
\affiliation{Instituto de F\'{\i}sica, Universidade de S\~ao Paulo, CP 66318, 05315-970 S\~aoPaulo, SP, Brazil}
\author{S.L. Morelh\~ao}
\email{morelhao@if.usp.br}
\affiliation{Instituto de F\'{\i}sica, Universidade de S\~ao Paulo, CP 66318, 05315-970 S\~aoPaulo, SP, Brazil}

\date{\today}
\begin{abstract}

Ultra-precise lattice parameter measurements in single crystals are achievable, in principle, by 
X-ray multiple diffraction (XMD) experiments. Tiny sample misalignments have hindered the systematic 
usage of XMD in studies where accuracy is an important issue. In this work, theoretical basement 
and methods for correcting general misalignment errors are presented. As a practical demonstration, 
the induced strain of buried InAs quantum dots grown on GaAs (001) substrates is determined. Such 
a demonstration confirms the possibility to investigate epitaxial nanostructures via the strain field 
that they generate in the substrate crystalline lattice.

\end{abstract}



\maketitle

\section{Introduction}

For decades, X-ray multiple diffraction (XMD) phenomenon has been applied in material science for 
studying a broad range of subjects: in-plane crystalline perfection of surfaces and interfaces 
\cite{[1],[2]}; strain determination in single crystals caused by an external stimulus 
\cite{[3]}$^{-}$\cite{[6]}; the studies of habit modification of crystals due to the introduction 
of impurities species \cite{[7]}; direct observation of crystallographic phase transition in single 
crystals \cite{[8]}; physical measurements of the ¨invariant phase triplets¨ \cite{[9]}$^{-}$\cite{[13]}; 
two-dimensional materials \cite{[14]} or quasi-crystals \cite{[15], [16]} and even in the plasma 
studies and diagnostics \cite{[17], [18]}.

Among the reasons behind the success of the XMD in material science, one comes from the fact that XMD 
is the unique technique capable of providing information over several different directions within a 
crystalline lattice under the same diffraction geometry, \textit{i.e.} with incidence $\omega$ and detector 
$2\theta$ fixed angles. Azimuthally scanning a chosen reflection, also known as Renninger scanning, 
allows accurate unit-cell lattice-parameters determination. This information is very important 
for characterizing single crystals under extreme conditions, where the instrumental setup, crystal shape 
and/or limited choise of X-ray photon energies do not leave too many accessible reflections to be measured. 
For a clear example, in applying uniform-uniaxial electric fields, electric contacts have to be set on 
two crystal faces. The field effects over the three-dimensional lattice can be studied by Renninger 
scanning a reflection that has not been hindered by the contacts \cite{[3]}. Exploitation of the synchrotron 
radiation linear polarization to tune the relative strength of the multiple diffracted waves has 
significantly increased the number of suitable reflections to be taken for Renninger scanning \cite{[19]}. 
Moreover, it is also well known that this technique is the most accurate method for experimental 
lattice parameters measurements in single crystals \cite{[20], [21]}; the precision can easily achieve 
1/10000\AA, depending on the chosen XMD case.

High-precision lattice parameter measurements, absolute or relative (variational), are necessary in 
several research fields, such as determination of piezoelectric tensor, structural phase transition 
and dopping with low concentrations of impurities (typically parts per million) where the effects 
produced over the unit cell are very small.  In studies where precise lattice parameter measurements 
are desired, it is essencial to account for all possible systematic errors in the Renninger scanning. 
A source of significant systematic errors originates from the misalingment between the azimuthal rotation 
axis $\phi$-axis and the diffraction vector of the reflection to be scanned, named the primary 
reflection. Even in very good equipments and careful sample alignment procedures, residual tiny 
misalingments are always present. Since, in general, these misalignments can be smaller than 
$0.01^{\circ}$, they have been neglected so far and no further efforts were made to take them into account. 
However, as experimentally observed, even such small misalignment severely compromise the accuracy of 
the results. In this article, we present the theory to account for a general sample misalignments as 
well as the experimental procedure for correcting systematic errors in lattice-parameter measurements 
by Renninger scanning.

\section{Theory}

To excite multiple X-ray diffraction in crystals, the incident beam direction, wavevector $\bm{k}$, 
must fulfill two conditions summarized by

\begin{equation}
\bm{k}\cdot\textbf{P}=-\textbf{P}\cdot\textbf{P}/2
\label{eq1}
\end{equation}
and
\begin{equation}
\bm{k}\cdot\textbf{S}=-\textbf{S}\cdot\textbf{S}/2.
\label{eq2}
\end{equation}

\textbf{P} and \textbf{S} are the diffraction vectors of the primary and secondary reflections, 
respectively. Equations ~(\ref{eq1}) and (\ref{eq2}) stand for the Bragg cones of these reflections in the 
reciprocal space. In Renninger scanning, the primary reflection is kept excited, \textit{i.e.} Eq.~(\ref{eq1}) 
is fulfilled, while the crystal rotates around \textbf{P}; it means that the incident beam is parallel 
to the primary Bragg cone during the azimuthal crystal rotation. The XMD occurs when the 
secondary reflection is excited by the azimuthal rotation and therefore Eq.~(\ref{eq2}) is also fulfilled. 
Note that more than one secondary reflection can be excited (3, 4,... beam cases) at the same azimuthal 
position; Eq.~(\ref{eq2}) is valid for all simultaneously diffracting secondary reflections. 

The most well-known expressions for determining XMD positions in Renninger scans were obtained from 
Eq.~(\ref{eq2}) \cite{[22], [23]} assuming that the primary reflection is always aligned, \textit{i.e.} 
Eq.~(\ref{eq1}) is fulfilled for a complete 360$^o$ rotation around \textbf{P}. However, it is not feasible 
to account for sample misalignments using the above couple of equations. The reason is that 
Eq.~(\ref{eq1}) is not sensitive to misalignments out of the incidence plane of the primary reflection;  
in other words, when the misalignment between \textbf{P} and the the azimuthal rotation axis ($\phi$ axis) 
is taken into account by Eq.~(\ref{eq1}), it appears in Eq.~(\ref{eq2}) as a negligible second-order 
correction. Therefore, a third equation where the sensitivity to the three-dimensional misalignment 
of \textbf{P} is the same as in Eq.~(\ref{eq2}) is necessary. Note that at least two equations are 
required since the XMD condition depends on the incidence $\omega$ and azimuthal $\phi$ angles
\textit{i.e.} depends on two variables.

A third equation that is also fulfilled under XMD condition stands for the Bragg cone of the coupling 
reflection, whose diffraction vector is given by \textbf{C}. Since $\textbf{P}=\textbf{S}+\textbf{C}$, 
it is possible to replace \textbf{P} by $\textbf{S}+\textbf{C}$ in Eq.~(\ref{eq1}) to obtain

\begin{equation}
\bm{k}\cdot\textbf{C}=-\textbf{C}\cdot\textbf{C}/2-\textbf{C}\cdot\textbf{S}.
\label{eq3}
\end{equation}

In this equation, the $\Delta\omega$ and $\Delta\phi$ deviations of the incident beam direction 
$\bm{k}(\omega,\phi)$ have now about the same weight that they have in Eq.~(\ref{eq2}); it leads to 
a trivial system of linear equations for determining both deviations as a function of the sample 
misalignment. The dependence of the incident beam direction on these angular deviations can be taken 
into account by

\begin{equation}
\bm{k}\simeq\bm{k}_0+\Delta\omega\bm{k}_{\omega}+\Delta\phi\bm{k}_{\phi}
\label{eq4}
\end{equation}

where $\bm{k}_{\omega}=\partial\bm{k}/\partial\omega$ and $\bm{k}_{\phi}=\partial\bm{k}/\partial\phi$ 
are calculated at $\omega_0$ and $\phi_0$, the incidence and azimuthal angles for exciting the XMD in 
non-misaligned samples, \textit{i.e.} at $\bm{k}_0=\bm{k}(\omega_0,\phi_0)$. Any reciprocal vector \textbf{G} 
can also be written in terms of $\textbf{G}^0$, its position in non-misaligned samples, and a rotation 
matrix $R$ as $\textbf{G}=\textbf{G}^0 R$. Then, by using Eq.~(\ref{eq4}) into Eqs.~(\ref{eq2}) and 
(\ref{eq3}), $\Delta\omega$ and $\Delta\phi$ are determined by solving the following system of linear 
equations \cite{[24]}:

\begin{equation}
\left[\begin{array}{cc}
\bm{k}_{\omega}\cdot\textbf{S} & \bm{k}_{\phi}\cdot\textbf{S}\\
\bm{k}_{\omega}\cdot\textbf{C} & \bm{k}_{\phi}\cdot\textbf{C}\end{array}\right]
\left[\begin{array}{c}
\Delta\omega\\
\Delta\phi\end{array}\right]=-
\left[\begin{array}{c}
(\textbf{S}/2+\bm{k}_0)\cdot\textbf{S}\\
(\textbf{C}/2+\textbf{S}+\bm{k}_0)\cdot\textbf{C}\end{array}\right].
\label{eq5}
\end{equation}

A common reference system is required to describe all vectors in the above equations. 
Here, the $(\hat{x},~\hat{y},~\hat{z})$ goniometer system shown in Fig. 1 will be used. $\hat{z}$ 
is along the $\phi$ axis, and $\hat{x}$ is an arbitrary reference direction chosen for $\phi=0$. 
For sake of simplicity, the vectors are represented by the $1 \times 3$ matrices of their components, 
for instance

\begin{figure}
\includegraphics[width=3.2in]{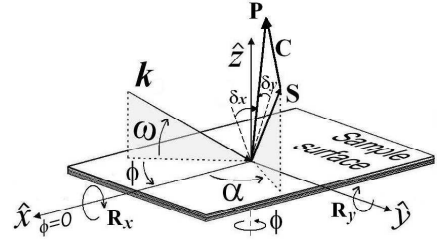}
\caption{\label{fig1} Three-beam X-ray diffraction in misaligned crystals where the primary diffraction 
vector \textbf{P} is not perfectly aligned along the azimuthal scanning axis ($\phi$ axis). The 
rotation matrices $R_x(\delta_x)$ and $R_y(\delta_y)$ account for the misalignment of the secondary 
and coupling diffraction vectors with respect to the $(\hat{x},\>\hat{y},\>\hat{z})$ goniometer 
reference system. $\bm{k}$ is the wavevector of the incident radiation.}
\end{figure}

\begin{subequations}
\begin{equation}
\bm{k}=-\lambda^{-1}(\cos\omega\cos\phi,~\cos\omega\sin\phi,~\sin\omega),
\label{eq6a}
\end{equation}
\begin{equation}
\bm{k}_{\omega}=-\lambda^{-1}(-\sin\omega_0\cos\phi_0,~-\sin\omega_0\sin\phi_0,~\cos\omega_0),
\label{eq6b}
\end{equation}
\begin{equation}
\bm{k}_{\phi}=-\lambda^{-1}(-\cos\omega_0\sin\phi_0,~\cos\omega_0\cos\phi_0,~0),
\label{eq6c}
\end{equation}
and
\begin{equation}
\textbf{G}^0=({\rm G}_x,~{\rm G}_y,~{\rm G}_z).
\label{eq6d}
\end{equation}
\end{subequations}

A general sample misalignment is taken into account by two rotation matrices $R_x(\delta_x)$ and 
$R_y(\delta_y)$ around the $\hat{x}$ and $\hat{y}$ directions (Fig. 1), and the total sample 
misalignment, for $\delta_{x,y}\ll1$, is 

\begin{equation}
R = R_y R_x \approx R_x R_y \approx 
\left[\begin{array}{ccc}
1 & 0 & -\delta_y\\
0 & 1 & -\delta_x\\
\delta_y & \delta_x & 1\end{array}\right].
\label{eq7}
\end{equation}

Therefore, 

\begin{equation}
\Delta\textbf{G}=\textbf{G}-\textbf{G}^0=({\rm G}_z\delta_y,~{\rm G}_z\delta_x,~-{\rm G}_x\delta_y-{\rm G}_y\delta_x)
\label{eq8}
\end{equation}

corresponds to the displacement of any diffraction vector due to crystal misalignment. 

The incident beam direction for non-misaligned samples $\bm{k}_0=\bm{k}(\omega_0,\phi_0)$ is calculated 
by Eqs.~(\ref{eq1}) and (\ref{eq2}), which provide $\omega_0$ and $\phi_0$ since

\begin{subequations}
\begin{equation}
\sin\omega_0=\lambda|\textbf{P}|^2/2
\label{eq9a}
\end{equation}
and
\begin{equation}
\cos(\phi_0-\alpha)=\cos\beta=\frac{\lambda|\textbf{S}|^2/2-{\rm S}_z\sin\omega_0}{{\rm S}_{xy}\cos\omega_0}.
\label{eq9b}
\end{equation}
\label{eq9}
\end{subequations}

where $\textbf{S}_x=\textbf{S}_{xy}\cos\alpha=\hat{x}\cdot\textbf{S}^0$ and 
$\textbf{S}_y=\textbf{S}_{xy}\sin\alpha=\hat{y}\cdot\textbf{S}^0$. 

According to Eq.~(\ref{eq9b}) there are two azimuthal positions, commonly called 'out-in' and 'in-out', 
where the same secondary reflection is excited: $\phi_1=\alpha-\beta$ and $\phi_2=\alpha+\beta$. 
Precise lattice-parameter determination procedures are based on the experimental measurements of both 
positions. They provide $2\beta_{exp}=\phi_{2,exp}-\phi_{1,exp}$ for the calculation of the unit cell 
parameters from Eqs.~(\ref{eq9}), as described elsewhere (see for instance reference \cite{[20]}). 
The sample misalignment is measured by rocking curves of the primary reflection at $\phi=0$, 
$90^{\circ}$, $180^{\circ}$, and $270^{\circ}$. If $\omega_0$, $\omega_{90}$, $\omega_{180}$ 
and $\omega_{270}$ are the respective rocking curves peak positions, 

\begin{equation}
\delta_x=(\omega_{270}-\omega_{90})/2\textrm{~~~and~~~}
\delta_y=(\omega_{180}-\omega_0)/2.
\label{eq10}
\end{equation}

The $\Delta\phi_n$ corrections in the 'out-in' ($n=1$) and 'in-out' ($n=2$) azimuthal positions are 
then obtained from Eq.~(\ref{eq5}), and they can be used either to refine the $2\beta_{exp}$ value 
according to

\begin{equation}
2\beta_{exp}=(\phi_{2,exp}-\Delta\phi_2)-(\phi_{1,exp}-\Delta\phi_1)
\label{eq11}
\end{equation}

or to estimate misalignment effects on different secondary reflections, for instance those providing some
useful information about the crystalline structure of the sample. To demonstrate this in practice, the 
in-plane strain of a GaAs (001) substrate induced by InAs quantum dots (QDs) is inspected in the following 
sections of this article.

\section{Experimental}

Data collection has been carried out at the Brazilian Synchrotron Light Laboratory (LNLS) D12A (XRD-1) 
beam-line, with the polarimeter-like diffractomer \cite{[25]}. The smallest step size is $0.0004^{\circ}$ 
in both $\omega$ and $\phi$ axes. The wavelength value, $\lambda=1.330234$\AA  ~from a Si (111) 
double-crystal monochromator, was measured by rocking curves the 111 and 333 Silicon reflections. The 
incidence plane was set to the vertical position, scattering upwards ($\sigma$-polarization). The sample 
used in this study is a comercial GaAs (001) substrate with InAs quantum dots (QD) grown by molecular 
beam epitaxy (MBE) using a rate of 0.009 monolayer per second (mL/s). A $300$\AA ~thick GaAs cap layer
was grown on top of the QDs.

\begin{figure} 
\includegraphics[width=1.6in]{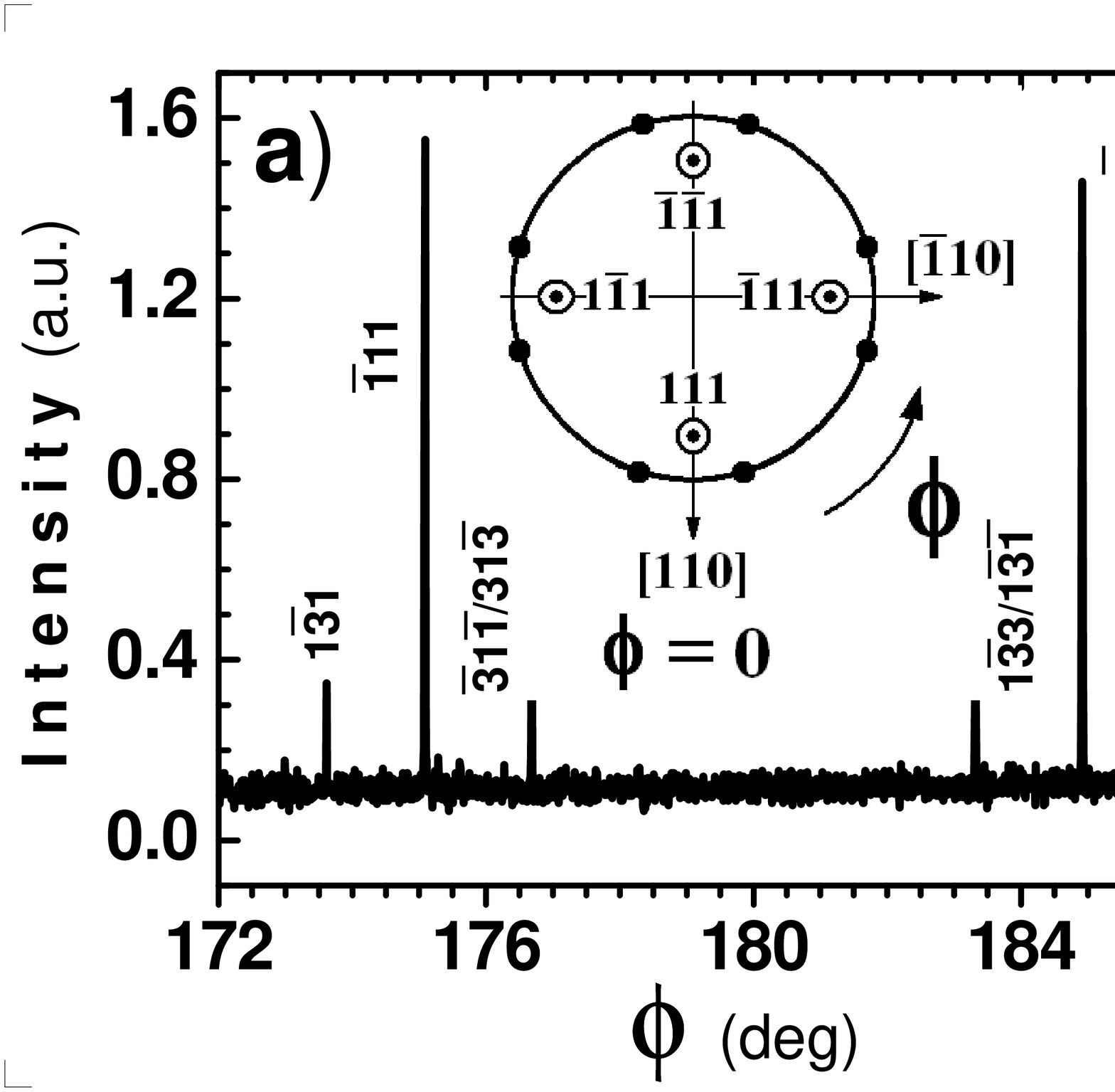}
\includegraphics[width=1.6in]{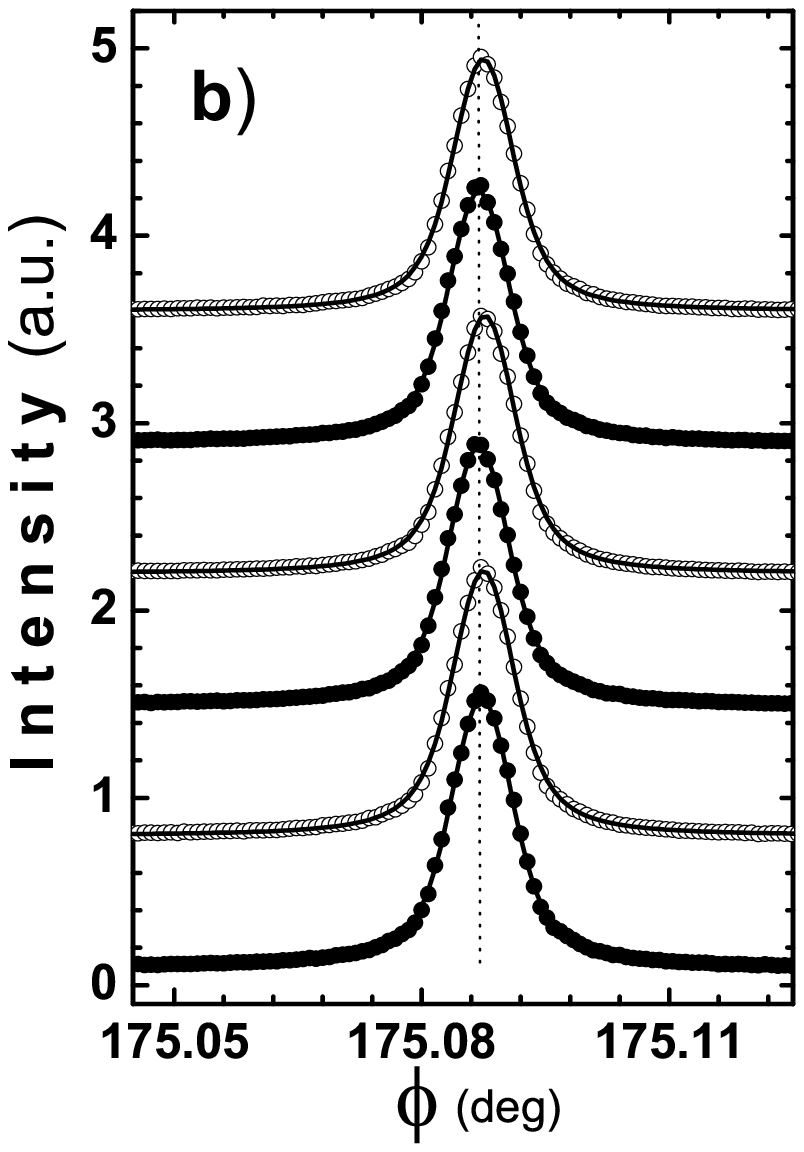}
\caption{\label{fig2} (a) Renninger scans of the 002 GaAs reflection carried out with X-ray wavelength 
$\lambda=1.330234$\AA. The highest peaks at $175.08^{\circ}$ and $184.92^{\circ}$ correspond to the 
excitation of the $\bar{1}11$ ('in-out') and $1\bar{1}1$ ('out-in') secondary reflections, respectively. 
The others secondary reflections shown in the diagram are also indexed. In the inset, $[110]$ is the reference 
direction, $\phi=0$, and the vectors coresponding to the $[111]$, $[\bar{1}11]$, $[\bar{1}\bar{1}1]$, and $[1\bar{1}1]$ 
directions are pointing outward the paper. The eight closed circles symbols indice the azimuthal $\phi$ positions 
of all secondary reflections and their values are shown in Table 1. (b) Several high-resolutions $\phi$ scans 
measured around $\phi = 175.08^{\circ}$ position (solid circle) and after a $360^{\circ}$ $\phi$ rotation (open 
circles). The peak positions have been determined by fitting the experimental intensity curves (open/solid circles) 
with lorentzian-gaussian convolution curves (solid lines), as explained in the text. The maximum shift observed 
in the peak position is $\phi_{max}-\phi_{min}=0.00075^{\circ}$.}
\end{figure}

\section{Results and discussions}

Table~\ref{tab1} shows the 'out-in' ($\phi_1$) and 'in-out' ($\phi_2$) azimuthal XMD peaks positions 
for the $111$, $\bar{1}11$, $\bar{1}\bar{1}1$, and $1\bar{1}1$ secondary reflections in the Renninger 
scanning of the 002 GaAs reflection. The peak positions were determined by fitting the intensity data 
with lorentzian-gaussian convolution curves: a lorentzian standing for the intensity profile 
(FWHM = 0.0048$^\circ$) of the peaks while the instrumental broadenig was taken into account by a 
gaussian function (FWHM = 0.0060$^\circ$). A short portion of the Renninger scan, around 
$\phi=180^{\circ}$, is shown in Fig. 2a; the $[110]$ direction was taken as reference for $\phi=0$. 
To assure that the data are free of instrumental errors, the $\phi$ scan of a chosen XMD peak, 
in this case the peak at $\phi=175.08^{\circ}$, was repeated after measuring the other peaks without 
changing the $\phi$ axis rotation sense, \textit{i.e.} after a $360^{\circ}$ $\phi$ rotation, see Fig.(2)b. 
Each $\phi$ scan has been performed at the (002) rocking-curve maximum which was carried out about 
$0.5^{\circ}$ before the XMD peak positions. 

The rocking curves used to characterize the sample misalignment are shown in Fig. 3; they provide 
$\delta_x\simeq0.0009^{\circ}$ and $\delta_y\simeq0.0067^{\circ}$. However, in general, for small sample 
misalignments Eq.~(\ref{eq5}) can be linearized by numerical derivation, and the corrections of the XMD 
$\phi$ positions can be written as $\Delta\phi_n={\rm A}_n\delta_x+{\rm B}_n\delta_y$. For the particular 
set of secondary reflections investigated here, the values of ${\rm A}_n$ and ${\rm B}_n$ coefficients 
are given in Table~\ref{tab1} (last two rows). By replacing these values into Eq.~(\ref{eq11}), it is 
possible to demonstrate that

\begin{subequations}
\begin{equation}
\bar{\beta}_{[110]}=(\beta_{111}+\beta_{\bar{1}\bar{1}1})/2
\label{eq12a}
\end{equation}
and 
\begin{equation}
\bar{\beta}_{[\bar{1}10]}=(\beta_{\bar{1}11}+\beta_{1\bar{1}1})/2
\label{eq12b}
\end{equation}
\label{eq12}
\end{subequations}

do not depend on $\delta_x$ and $\delta_y$. In other words, the average $\beta_{exp}$ for the secondary 
reflections with in-plane components in the $[110]$ and $[\bar{1}10]$ orthogonal directions, 
$\bar{\beta}_{[110]}$ and $\bar{\beta}_{[\bar{1}10]}$, are misalignment-free experimental values, and 
therefore, very useful for precise lattice parameter measurements in both orthogonal directions.

\begin{figure} 
\includegraphics[width=3.2in]{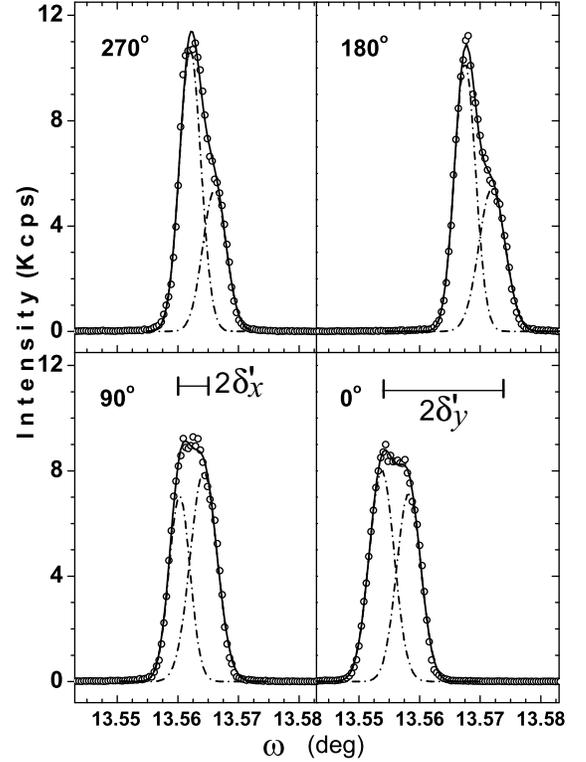}
\caption{\label{fig3} Experimental (open circles) and simulated (solid lines) rocking curves,
$\omega$-scans, of the 002 primary reflection measured at $\phi=0$, $90^{\circ}$, $180^{\circ}$ and
$270^{\circ}$, as shown aside of each scan. These scans were simulated by using two gaussian curves
(dashed lines) whose splitting is about $\Delta\omega=-0.00422(24)^{\circ}$. It provides $\Delta c/c =
3.0(2) \times 10^{-4}$. The misalignment angles calculated from the gaussians at lower $\omega$ values
are $\delta_x\simeq0.0009^{\circ}$ and $\delta_y\simeq0.0069^{\circ}$, while 
$\delta_x^{\prime}~(=0.0024^{\circ})$ and $\delta_y^{\prime}~(=0.0103^{\circ})$ were obtained from the
$\bar{\beta}$ values in Table~\ref{tab1}.}
\end{figure}

We assume a tetragonal lattice distortion where the substrate in-plane lattice parameter is given by 
$a=b=(1-\nu)a_0$ while, to preserve the unit cell volume, $c=(1+2\nu)a_0$. For $a_0=5.6534$\AA, 
$|\textbf{P}|=2/c$, $\textbf{S}^0=(h/a,~k/b,~1/c)$ (in the crystal reciprocal space) and $h,k=\pm1$,
Eq.~(\ref{eq9}) provides 

\begin{equation}
\beta(\nu)=85.08984^{\circ}-33.85^{\circ}\nu.
\label{eq13}
\end{equation}

Note that $|\textbf{S}|$=$|\textbf{S}^0|$ is invariant under any rotation, but their components are not. 
Therefore, ${\rm S}_x$, ${\rm S}_y$ and ${\rm S}_z$ in Eq.~(\ref{eq9b}) must be calculated for 
$\hat{x}=(1,~1,~0)/\sqrt{2}$, $\hat{y}=(\bar{1},~1,~0)/\sqrt{2}$ and $\hat{z}=(0,~0,~1)$, \textit{i.e.}, 
${\rm S}_x=(h+k)/(a\sqrt{2})$, ${\rm S}_y=(-h+k)/(a\sqrt{2})$ and ${\rm S}_z=1/c$.

\begin{table*} 
\caption{\label{tab1}Azimuthal positions of the $111$, $\bar{1}\bar{1}1$, $\bar{1}11$, and
$1\bar{1}1$ secondary reflections in the Renninger scanning of the 002 GaAs primary reflection.
Each position was measured three times (rows 1, 2 and 3) as explained in the text,
$\bar{\phi}=(\phi_{max}+\phi_{min})/2$, $\varepsilon=(\phi_{max}-\phi_{min})/2$,
$\beta_{exp}=(\bar{\phi}_2-\bar{\phi}_1)/2$, $\bar{\beta}$ are the average misalignment-free
values, $\nu$ stands for the unit-cell tetragonal distortion (Eq.~(\ref{eq13})) and
$\beta_{mis}=\bar{\beta}+(\Delta\phi_2-\Delta\phi_1)/2$ (Eq.~(\ref{eq11})), where
$\Delta\phi_n={\rm A}_n\delta_x^{\prime}+{\rm B}_n\delta_y^{\prime}$,
$\delta_x^{\prime}=0.0024^{\circ}$, and $\delta_y^{\prime}=0.0103^{\circ}$. ${\rm A}_n$ and 
${\rm B}_n$ were estimated by numerical derivation of Eq.~(\ref{eq5}). Angular values are given in 
degrees.}

\begin{ruledtabular} 
\begin{tabular}{ccccccccc}
&\multicolumn{2}{c}{$111$}&\multicolumn{2}{c}{$\bar{1}\bar{1}1$}&\multicolumn{2}{c}{$\bar{1}11$}&\multicolumn{2}{c}{$1\bar{1}1$}\\
 & $\phi_1$ & $\phi_2$ & $\phi_1$ & $\phi_2$ & $\phi_1$ & $\phi_2$ & $\phi_1$ & $\phi_2$\\ \hline
1&-85.07750&85.08599&94.91867&265.09121&4.92085&175.08692&184.92072&355.08935\\ 
2&-85.07716&85.08575&94.91853&265.09141&4.92102&175.08688&184.92061&355.08881\\
3&-85.07683&85.08548&94.91819&265.09209&4.92105&175.08730&184.92064&355.08880\\
$\bar{\phi}$&-85.077165&85.085735&94.91843&265.09165&4.92095&175.08709&184.920665&355.089075\\
$\varepsilon$&$\pm$0.000335&$\pm$0.000255&$\pm$0.00024&$\pm$0.00044&$\pm$0.00010&$\pm$0.00021&$\pm$0.000055&$\pm$0.000275\\
$\beta_{exp}$&\multicolumn{2}{c}{85.081450$\pm$0.000295}&\multicolumn{2}{c}{85.08661$\pm$0.00034}&\multicolumn{2}{c}{85.083070$\pm$0.000155}&\multicolumn{2}{c}{85.084205$\pm$0.000165}\\
$\bar{\beta}$&\multicolumn{4}{c}{85.08403$\pm$0.00032}&\multicolumn{4}{c}{85.08364$\pm$0.00016}\\
$\nu$&\multicolumn{4}{c}{$(1.716\pm0.096)\times 10^{-4}$}&\multicolumn{4}{c}{$(1.832\pm0.048)\times 10^{-4}$}\\
$\beta_{mis}$&\multicolumn{2}{c}{85.0815}&\multicolumn{2}{c}{85.0865}&\multicolumn{2}{c}{85.0831}&\multicolumn{2}{c}{85.0842}\\
${\rm A}_n$&0.0207&0.0207&-0.0207&-0.0207&0.2412&-0.2412&-0.2413&0.2413\\
${\rm B}_n$&0.2412&-0.2412&-0.2413&0.2413&-0.0207&-0.0207&0.0207&0.0207\\
\end{tabular}
\end{ruledtabular}
\end{table*}

\begin{figure} 
\includegraphics[width=3.2in]{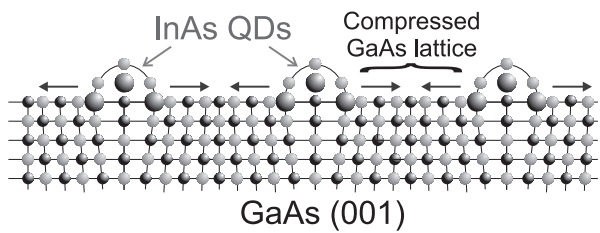}
\caption{\label{fig4} Schematic illustration of the compressive stress generated by InAs QDs on GaAs 
substrate. There is also a 300\AA ~thick GaAs cap layer over the QDs (not shown in the figure).}
\end{figure}

In the 002 Renninger scan of (001) samples, $hk1$ secondary reflections give rise to 
special XMD cases, named Bragg-surface diffractions, where the secondary-beam directions are very close 
to surface in-plane directions. The extreme asymmetry of such secondary reflections limits the secondary 
wavefield penetration depth to less than $1\>\mu m$. Experimentally, this value has been measured by 
reciprocal-space mapping of the 002 rod in GaAs (001) crystals undergoing Bragg-surface diffraction.
It has been demonstrated that about 80\% of the diffraction intensity, for X-ray photons of 8 keV, 
is attenuated in the first $0.3\>\mu m$ bellow the surface~\cite{[26]}. According to this fact, the 
observed tetragonal deformation would correspond to an average value of a shallow layer just below the 
surface where the induced strain due to the InAs QD is significant. Although the InAs epitaxial growth 
generates an expansive stress in the substrate lattice under the QDs, the adjacent regions are 
compressed as schematically illustrated in Fig.~\ref{fig4}.

Differences in the tetragonal distortion along the $[110]$ and $[\bar{1}10]$ directions, \textit{i.e.} 
$\bar{\beta}_{[110]}\neq\bar{\beta}_{[\bar{1}10]}$, would yield a unit cell sligthly twisted near the 
substrate surface. For instance, tiny variations in the $\gamma$ angle, between the $\bm{a}$ and $\bm{b}$ 
lattice vectors, can be estimated by numerical derivation of Eq.~(\ref{eq9}). In this case, it leads to 

\begin{equation}
\bar{\beta}_{[\bar{1}10]}-\bar{\beta}_{[110]}=(C_{\bar{1}10}-C_{110})\Delta\gamma
\label{eq14}
\end{equation}

where $C_{\bar{1}10}=0.1276$ and $C_{110}=-0.1302$, and then to $\gamma=89.9985^{\circ}\pm0.0018^{\circ}$.

Finally, the agreement between the $\delta_x$ and $\delta_y$ and $\delta_x^{\prime}$ and $\delta_y^{\prime}$ 
values is better when rocking curves are measured with higher definition in their maximum peak position.

\section{Conclusions}

In summary, a method that corrects any sample misalignment and allows the determination of highly accurate 
single-crystals lattice parameters or their variation was presented. The advantage of this method is in the
measurement of rocking curves in orthogonal positions. By combining those rocking curves and measurements of 
a XMD secondary reflection family, having the same $\beta$ value and measured in their "in-out" and "out-in" 
azimuthal positions, the effects of a residual sample misalignment is eliminated and any systematic error 
is averaged out. Therefore, a precision better than 1/10000\AA ~can be easily achieved.

As the method presented here can determine very small single crystal lattice deformations, it is an excellent 
tool for studying epitaxial nanostructures grown on top of crystalline substrates. The strain field produced 
by the nanostructures in the substrate lattice can be determined even when those nanostructures are buried 
by cap layers. As an application, QDs grown on GaAs (001) were analyzed and the results showed that the GaAs 
lattice is tetragonally distorted and slightly twisted near the substrate surface.

The method however, is not restricted to the particular geometry of the QDs system studied in this work where, 
due to the primary diffraction vector high symmetry, a family of eight secondary reflections could be measured. 
When the system symmetry under study is lower, it is only important to obtain rocking-curves profiles with 
well-defined peak positions.

\begin{acknowledgments}

This work was supported by the Brazilian founding agencies FAPESP (grant numbers 02/10185-3 and 02/10387-5), CNPq (proc. 
number 301617/95-3 and 150144/03-2) and LNLS (under proposal number D12A-XRD1 2490/03). 

\end{acknowledgments}


\begin{thebibliography}{26}.

\bibitem{[1]} S. L. Morelh\~ao and L. P. Cardoso, J. Appl. Cryst. \textbf{29}(4), 446-456 (1996).
\bibitem{[2]} M. A. Hayashi, S. L. Morelh\~ao, L. H. Avanci, L. P. Cardoso, J. M. Sasaki, L. C. Kretly
and S.-L. Chang, Appl. Phys. Lett. \textbf{71}(18), 2614-2616 (1997).
\bibitem{[3]} L. H. Avanci, L. P. Cardoso, S. E. Girdwood, D. Pugh, J. N. Sherwood and K. J. Roberts, Phys. Rev. Lett. 
\textbf{81}, 5426-5429 (1998).
\bibitem{[4]} L. H. Avanci, L. P. Cardoso, J. M. Sasaki, S. E. Girdwood, K. J. Roberts, D. Pugh and J. N. Sherwood, Phys. 
Rev. B \textbf{61}, 6507-6514 (2000).
\bibitem{[5]} J. M. A. Almeida, M. A. R. Miranda, C. M. R. Rem\'edios, F. E. A. Melo, P. T. C. Freire, J. M. Sasaki, L. P. Cardoso, 
A. O. dos Santos and S. Kycia, J. Appl. Cryst. \textbf{36}(6), 1348-1351 (2003).
\bibitem{[6]} A. O. dos Santos, L. P. Cardoso, J. M. Sasaki, M. A. R. Miranda and F. E. A. Melo, J. Phys.-Condensed Matter 
\textbf{15}(46), 7835-7842 (2003).
\bibitem{[7]} X. Lai, K. J. Roberts, L. H. Avanci, L. P. Cardoso and J. M. Sasaki, J. Appl. Cryst. \textbf{36}(5), 1230-1235 (2003).
\bibitem{[8]} S.-L. Chang, Appl. Phys. Lett. \textbf{37}, 819-821 (1980).
\bibitem{[9]} R. Colella, Acta Cryst. Sect. A \textbf{30}, 413-423 (1974).
\bibitem{[10]} H. J. Juretschke, Phys. Rev. Lett. \textbf{48}, 1487-1489 (1982).
\bibitem{[11]} S.-L. Chang, Int. J. Mod. Phys. B \textbf{6}, 2987-3020 (1992) and references therein.
\bibitem{[12]} E. Weckert and K. Hümmer, Acta Crystallogr. Sect. A \textbf{53}, 108-143 (1997) and references therein.
\bibitem{[13]} S. L. Morelh\~ao and S. Kycia, Phys. Rev. Lett. \textbf{89}(1), Art. No. 015501 (2002).
\bibitem{[14]} C.-H. Du, M.-T. Tang, Y. P. Stetsko, Y.-R. Lee, C.-W. Wang, C.-W. Cheng and S.-L. Chang, Acta Cryst. 
A \textbf{60}, 209-213 (2004).
\bibitem{[15]} Y. Zhang, R. Colella, Q. Shen and S. W. Kycia, Acta Cryst. Section A \textbf{54}(4), 411-415 (1998).
\bibitem{[16]} Q. Shen, S. Kycia and I. Dobrianov, Acta Cryst. Section A \textbf{56}(3), 268-279 (2000).
\bibitem{[17]} B. S. Fraenkel, Appl. Phys. Lett. \textbf{36}, 341-343 (1980).
\bibitem{[18]} B. S. Fraenkel, Appl. Phys. Lett. \textbf{41}, 234-236 (1982).
\bibitem{[19]} S. L. Morelh\~ao and L. H. Avanci, Acta Cryst. Section A \textbf{57}(2), 192-196 (2001
\bibitem{[20]} B. Post, J. Appl. Cryst. \textbf{8}, 452-456 (1975).
\bibitem{[21]} T. Hom, W. Kiszenick and B. Post, J. Appl. Cryst. \textbf{8}, 457-458 (1975).
\bibitem{[22]} H. Cole, F. W. Chambers and H. M. Dunn, Acta Cryst. \textbf{15}, 138-144 (1962).
\bibitem{[23]} S. Caticha Ellis, Jpn. J. Appl. Phys. \textbf{14}(5), 603-611 (1975).
\bibitem{[24]} S. L. Morelh\~ao, A. A. Quivy and J. H\"artwig, Microel. Journal \textbf{34}, 695-699 (2003).
\bibitem{[25]} S. L. Morelh\~ao, J. Sync. Rad. \textbf{10}, 236-241 (2003).
\bibitem{[26]} S. L. Morelh\~ao and E. Abramof, J. Appl. Cryst. \textbf{33}, 871-877 (1999).
\end{thebibliography}
\end{document}